\documentclass[a4paper,fleqn,usenatbib]{mnras}
\usepackage{newtxtext,newtxmath}
\usepackage[T1]{fontenc}
\usepackage{ae,aecompl}
\usepackage{graphicx}	
\usepackage{amsmath}	
\usepackage{amssymb}	
\usepackage{pdflscape}

\newcommand{\lbolint}{\ifmmode L(IR;X-rays) \else $L${\it (IR; X-rays)}\fi}
\newcommand{\ledd}{\ifmmode L_{Edd}\else $L_{Edd}$\fi}
\newcommand{\lbol}{\ifmmode L_{BOL}\else $L_{BOL}$\fi}
\newcommand{\Mdot}{\ifmmode \dot{M} \else $\dot{M}$\fi}

\title[]{The Infrared Luminosity Function of AKARI 90$\mu$m Galaxies in the Local Universe}

\author[E. Kilerci Eser and T. Goto]{
Ece Kilerci Eser,$^{1}$\thanks{E-mail: ecekilerci@phys.nthu.edu.tw (EKE)}
T. Goto,$^{1}$
\\
$^{1}$Institute of Astronomy , National Tsing Hua University, No. 101, Section 2, Kuang-Fu Road, Hsinchu, 30013, Taiwan\\
}

\date{Accepted XXX. Received YYY; in original form ZZZ}

\pubyear{2017}

\begin{document}
\label{firstpage}
\pagerange{\pageref{firstpage}--\pageref{lastpage}}
\maketitle

\begin{abstract}
 Local infrared (IR) luminosity functions (LFs) are necessary benchmarks for high-redshift IR galaxy evolution studies.
 Any accurate IR LF evolution studies require accordingly accurate local IR LFs.

 We present infrared galaxy LFs at redshifts of $z \leq 0.3$ from \textit{AKARI} space telescope, which performed an all-sky survey in six IR 
 bands (9, 18, 65, 90, 140 and 160 $\mu$m) with 10 times better sensitivity than its precursor IRAS.
 Availability of 160 $\mu$m filter is critically important in accurately measuring total IR luminosity of galaxies, covering across the peak of the dust emission. 
  By combining data from Wide-field Infrared Survey Explorer (\textit{WISE}), 
 Sloan Digital Sky Survey (SDSS) Data Release 13 (DR13), 6-degree Field Galaxy Survey (6dFGS) and the 2MASS Redshift Survey (2MRS), 
 we created a sample of 15,638 local IR galaxies with spectroscopic redshifts, factor of 7 larger compared to previously studied \textit{AKARI}-SDSS sample.
 After carefully correcting for volume effects in both IR and optical,
 the obtained IR LFs agree well with previous studies, but comes with much smaller errors.  
 Measured local IR luminosity density is $\Omega_{IR}= 1.19\pm0.05 \times 10^{8} L_{\sun}$ Mpc$^{-3}$. 
 The contributions from luminous infrared galaxies and ultra luminous infrared galaxies to $\Omega_{IR}$ are very small, 9.3 per cent and 0.9 per cent, respectively.
 There exists no future all sky survey in far-infrared wavelengths in the foreseeable future.
 The IR LFs obtained in this work will therefore remain an important benchmark for high-redshift studies for decades.
\end{abstract}




\section{INTRODUCTION}\label{introduction}

Luminosity function (LF) represents the number density of galaxies as a function of luminosity. 
Luminosity functions are crucial cosmological observables to understand galaxy evolution 
and structure formation \citep[e.g.,][]{Benson2003,Croton2006,Bower2010,Trayford2015,Steinhardt2016} at different redshifts.
The galaxy luminosity functions  have been measured at optical \citep[e.g.,][]{Blanton2001,Blanton2003,Madgwick2002,Bell2003,Montero-Dorta2009,Cool2012}, 
ultraviolet (UV) \citep[e.g.,][]{Reddy2008,Bowler2014,Heinis2013,Finkelstein2015}, 
near-infrared (i.e., $K$-band)  \citep[e.g.,][]{Kochanek2001,Bell2003,Eke2005,Jones2006}, 
mid-infrared (i.e., 3.6$\ge \lambda \leq $40 $\mu$m) \citep[e.g.,][]{LeFloch2005,Perez-Gonzalez2005,Babbedge2006,Caputi2007,Dai2009,Goto2010,Fu2010,Rodighiero2010,Magnelli2011,Wu2011}, 
far-infrared (i.e., $\lambda > 40 \mu$m) \citep[e.g.,][]{Huynh2007,Sedgwick2011,Patel2013,Gruppioni2013,Marchetti2016}, submillimeter \citep[e.g.,][]{Vaccari2010,Negrello2013} wavelengths and 
from total infrared luminosities ($L_{IR}$, integrated over 8$-$1000 $\mu$m) 
\citep[e.g.,][]{LeFloch2005,Goto2010,Rodighiero2010,Goto2011a,Goto2011,Magnelli2011,Sargent2012,Patel2013,Heinis2013,Magnelli2013,Gruppioni2013,Marchetti2016}. 
It is evident from all of these studies that in order to constrain models of galaxy formation and evolution  
we need accurate galaxy LFs at different wavelengths in the local and distant Universe. 

Infrared (IR) emission is a practical tool to determine star formation rates of galaxies. 
Especially, when the star formation is embedded in dust, the star-formation activity of galaxies can only be observed in the infrared.
Therefore, FIR emission is very important to reveal the star formation in the Universe that is hidden by dust \citep[e.g.,][]{Rowan-Robinson2001,Lagache2003,Goto2010}. 
Specifically the evolution of the IR luminosity function is one of the most useful tools to explore the evolution of star formation in galaxies. 

The all-sky survey performed by the \textit{Infrared Astronomical Satellite} \citep[\textit{IRAS},][]{Neugebauer1984} revealed 
the numbers and properties of the IR galaxies including the IR LFs in the local Universe \citep[][]{Soifer1987,Saunders1990,Sanders2003}. 
The brighter IR galaxies, luminous and ultraluminous infrared galaxies are found to be rare local objects; however, they are more common at higher redshifts.  
Infrared satellites including the \textit{Infrared Space Observatory} \citep[\textit{ISO},][]{Kessler1996} and \textit{Spitzer} Space Telescope \citep[][]{Werner2004} carried out deep cosmological surveys. 
These observations imply a strong evolution (between $z= 0 - 2$) of the LF, particularly at the bright end \citep[e.g.,][]{Elbaz1999,LeFloch2005,Perez-Gonzalez2005,Caputi2007,Magnelli2009,Rodighiero2010}.  
Recently, Wide-field Infrared Survey Explorer \citep[\textit{WISE},][]{Wright2010} provided another opportunity to measure the 
IR luminosity function up to $z\sim 1$ \citep[e.g.,][]{Lake2017}.
Our knowledge of the high-redshift IR galaxies increased with the \textit{Herschel} Space Observatory \citep[][]{Pilbratt2010}. 
The observed SEDs from ultraviolet to submillimetre enable to determine IR LFs of high-redshift (beyond $z=3$) galaxies observed by 
\textit{Herschel} \citep[e.g.,][]{Gruppioni2010,Eales2010,Gruppioni2013,Casey2012}.
Observations based on galaxies at different redshifts show a dramatic redshift evolution, both in IR LF and IR luminosity density. 
Since all high-redshift studies need to be compared with local ones, the investigation of the redshift evolution highly depends on the local galaxies. 

\textit{AKARI} performed an all-sky survey with better spatial resolution, sensitivity and FIR coverage (between 50$-$180 $\mu$m) compared to \textit{IRAS}.  
The advent of \textit{AKARI} has provided a unique FIR data set. 
\textit{AKARI} has four FIR filters centred at 65$\mu$m, 90 $\mu$m, 140$\mu$m and 160 $\mu$m. 
Therefore, it has a great advantage to constrain the peak of the FIR SED around  90$\mu$m. 
\citet{Goto2011} used a larger sample of local IR galaxies than previous all-sky samples identified by \textit{IRAS} \citep{Sanders2003}. 
They use \textit{AKARI} all-sky survey photometry and SDSS DR-7 spectroscopic redshifts to  measure accurate $L_{IR}$. 
They also construct the LFs of 2357 local IR galaxies. 
Motivated by the results of \citet{Goto2011}, we intend to improve their investigation of local IR LF. 
By using \textit{AKARI} all-sky survey photometry, but matching with a much larger optical survey area in northern and southern sky, we establish the local IR LF. 
We have an optically and 90 $\mu$m limited sample with the advantage of 90 $\mu$m detections that can accurately constrain $L_{IR}$ from SED fitting based on spectroscopic redshifts.

In this work, we measure the local IR luminosity  function with the largest sample (15,638) of galaxies yet used for this purpose. 
We aim for the most accurate local IR LF which will be a reliable local benchmark for high-redshift studies especially in the era of new missions like 
Space Infrared Telescope for Cosmology and Astrophysics \citep[\textit{SPICA},][]{Nakagawa2012} and 
James Webb Space Telescope \citep[\textit{JWST},][]{Gardner2006}. 
The structure of this paper is as follows. 
In section \ref{S:data} we present the sample selection and data. 
We describe the total IR luminosity measurements and total IR LFs of the \textit{AKARI}$-$SDSS,  \textit{AKARI}$-$6dFGS 
and the combined \textit{AKARI}$-$SDSS$-$6dFGS$-$2MRS samples  in \S \ref{S:results}. 
We summarise our conclusions in \S \ref{S:conc}. 
We adopt a cosmology with $H_0=72$\,km\,s$^{-1}$\,Mpc$^{-1}$, $\Omega_\Lambda = 0.7$ and $\Omega_{\rm m}=0.3$ throughout. 
We use the base 10 logarithm.

\section{SAMPLE SELECTION and DATA} \label{S:data}

\subsection{AKARI$-$WISE Sample}\label{S:samples}

\textit{AKARI} is an infrared astronomy satellite \citep{Murakami2007} funded by the Institute of Space and Astronautical Science (ISAS) of the Japanese Aerospace Exploration Agency (JAXA) and launched in February 2006. \textit{AKARI} performed an all-sky survey in two  mid-IR bands at 9 and 18 $\mu$m, and four far-IR bands centred respectively at 65, 90, 140, and 160 $\mu$m.
The \textit{AKARI}/FIS all-sky survey bright source catalog version 2\footnote[1]{http://www.ir.isas.jaxa.jp/AKARI/Archive/Catalogues/FISBSCv2/} \citep[][in preparation]{Yamamura2017inprep} provides the positions and fluxes at 65, 90, 140 and 160$\mu$m for 918,056 sources (this is the sum of the \textit{main} and the \textit{supplemental catalogue}.). This catalog is the revised version of the 
\textit{AKARI}/FIS all-sky survey bright source catalog version 1\footnote[2]{http://www.ir.isas.jaxa.jp/AKARI/Observation/PSC/Public/RN/AKARI-FIS\_BSC\_V1\_RN.pdf} with better position and flux accuracy, detection completeness and reliability. 

The far-IR photometric measurements of \textit{AKARI}  are crucial to identify IR galaxies whose SEDs peak at 60$-$100$\mu$m. 
The goal of our study is to identify and study local IR galaxies. 
Therefore, we base our study on the \textit{AKARI}/FIS all-sky survey bright source catalog version 2. 
In order to have the most reliable far-IR measurements we select only the sources with $FQUAL(90\mu m) = 3$ and the \textit{AKARI} photometric fluxes of the 90$\mu$m (F(90$\mu$m)) greater than 0.5 Jy. 
Thus, our initial \textit{AKARI} sample has 540,978 sources. 
 
The \textit{AKARI}/IRC all-sky survey point source catalog version 1\footnote[3]{http://www.ir.isas.jaxa.jp/AKARI/Observation/PSC/Public/RN/AKARI-IRC\_PSC\_V1\_RN.pdf} provides mid-IR photometry at 9 and 18 $\mu$m for 870,973 sources. 
In order to include these mid-IR measurements in our data set we cross-match 
the initial \textit{AKARI} sample with the \textit{AKARI}/IRC all-sky survey point source catalog version 1 using a search radius of 20\arcsec. 
We find mid-IR measurements for 28,578 sources. 
 
Most of the sources (512,400) in our initial \textit{AKARI} sample do not have mid-IR measurements from \textit{AKARI}, therefore
we include \textit{WISE} data to have more sources with mid-IR photometry. 
The \textit{WISE} completed an all-sky survey in four mid-IR bands, $W1$, $W2$, $W3$, $W4$, centred at 3.4, 4.6, 12, and 23 $\mu$m, respectively. 
The AllWISE Source Catalog\footnote[4]{http://wise2.ipac.caltech.edu/docs/release/allwise/} provides positions, photometric measurements, quality information for over 747 million sources 
\citep{Cutri2013}. We match our initial \textit{AKARI} sample with AllWISE Source Catalog using a 20\arcsec search radius. 
We only include \textit{WISE} sources with zero $cc\_flags$ values ($cc\_flags='0000'$) to avoid objects with contaminated measurements (i.e., by diffraction spikes, bright sources, optical ghost images). 
We also require that sources have S/N greater than 2 in any of the 4 bands to eliminate upper limit values. 
We choose such a low S/N ratio mainly to avoid upper limits and to keep as many \textit{WISE} sources as possible in our sample, and also note that we already have source detection at 90 $\mu$m 
(in our final samples, sections \ref{S:AKARIcatalogs}, \ref{S:6df} and \ref{S:2mrs} the S/N in all 4 \textit{WISE} bands are between 2.0 and 141.3 with a median value of 46.5).   
Additionally we require  standard aperture measurement quality flag ($w1,2,3,4flg$) to be smaller than 32 in any of the 4 bands to exclude sources with upper limits. 
We eliminated the magnitude in any band with \textit{null} uncertainties (i.e., if `$wnsigmpro$' photometric measurement uncertainty is null it indicates an upper limit or no measurement, where $n$ refers to the band number); 
with high saturated pixel fraction (i.e., if the fraction of saturated pixels in any band is greater than 0.05); with high scattered moonlight contamination (i.e., `{\ttfamily moon$\_$lev}' $\geq 5$ ); 
with measurement quality flag greater than 1 (i.e., `$wnflg$' and `$wngflg$' indicate pixel saturation, confusion with other objects, unusable pixels or upper limits, where $n$ refers to the band number).
We also check the extended source flag (extended sources have $ext\_flg > 0$) and use the elliptical aperture magnitudes for these sources (`$wngmag$', $n$ refers to the band number).
Our \textit{AKARI}$-$\textit{WISE} sample consists of 208,459 sources. 
In the initial \textit{AKARI} sample the number of sources without \textit{WISE} measurements is 332,519; 62\% of  \textit{AKARI} sources do not have a \textit{WISE} detection. 

\subsection{Cross-Correlation with Optical Spectroscopic Redshift Catalogs}\label{S:cross}
In order to include most of the northern and southern sky, we search SDSS Data Release 13 \citep[DR 13;][]{Albareti2016}, the Six-degree Field Galaxy Survey \citep[][]{Jones2004,Jones2005,Jones2009} and the Two Micron All Sky Survey \citep[][]{Huchra2005,Huchra2012} public spectroscopic catalogs to find redshifts of the IR sources with optical counterparts. 
Since position accuracy of \textit{WISE} is more accurate than \textit{AKARI} for the cross-match between infrared and optical sources we use \textit{WISE} coordinates with a cross-match radius of 5\arcsec, where available. For \textit{AKARI} sources without \textit{WISE} positions we use \textit{AKARI} FIS coordinates and use a matching radius of 20\arcsec. 
 
\subsubsection{The AKARI$-$SDSS DR13 sample}\label{S:AKARIcatalogs}

The Sloan Digital Sky Survey (SDSS) is a ground-based imaging and spectroscopy survey started in 2000 \citep{York2000, Abazajian2009,Eisenstein2011}.  
The data release DR 13 of SDSS fourth phase \citep[SDSS-IV,][]{Blanton2017} covers 14,555 square degrees, more than one-third of the whole sky \citep{Albareti2016}. 
The \textit{specObj}\footnote[5]{https://data.sdss.org/sas/dr13/sdss/spectro/redux/} catalog lists positions ($PLUG\_RA, PLUG\_DEC$), best spectroscopic redshift ($z$) and spectral class ( as "GALAXY", "QSO" or "STAR"). Since our main goal is to study galaxies, we exclude sources with "STAR" classification and left with 3,400,232 sources in the \textit{specObj} catalog. 

We cross-match our initial \textit{AKARI} sample with SDSS-DR13 \textit{specObj} catalog. 
The \textit{AKARI}$-$SDSS sample has 6,181 sources, among these 5,786 are \textit{WISE} sources. 
We check the redshift quality flag, $ZWARNING$, given by SDSS. 6,019 sources have reliable secure measurements ($ZWARNING=0$) and 162 sources 
have uncertain redshift values $ZWARNING > 0$. The spectra of these 162 sources are visually inspected. 
Among this subset, 61 sources are excluded because of their unreliable spectra. 
We further restricted the \textit{AKARI}$-$SDSS sample to those galaxies with extinction corrected $r_{petro} \leq 17.7$ within the $0.02 \leq z \leq 0.3$ redshift range. 
We adopt the spectroscopic redshift completeness estimate given by \citep{Montero-Dorta2009} and use a lower redshift limit of 0.02. 
As a result, the \textit{AKARI}$-$SDSS sample consist of 4,705 local IR galaxies with reliable spectroscopic redshift and photometric measurements. 

\subsubsection{The AKARI$-$6dFGS sample}\label{S:6df}
The Six-degree Field Galaxy Survey \citep[6dFGS][]{Jones2004,Jones2005,Jones2009} covers almost the entire southern sky 10\textdegree~above the Galactic plane. 
The 6dF Galaxy Survey Redshift Catalogue Data Release 3 (final) catalogue\footnote[6]{http://www-wfau.roe.ac.uk/6dFGS/download.html} lists redshifts of 124,647 sources. 
 Among those we only keep 108,030 sources classified as a galaxy. 
We cross-match the initial \textit{AKARI} sample with the extracted 6dF catalog and obtain 7,547 sources; 6,654 are \textit{WISE} sources. 
For 8 sources the redshift comparison flag indicate there is a disagreement between SDSS and 6dFGS redshift values or between different 6dFGS measurements. Since these are not sole measurements we exclude these 8 sources from the 6dFGS sample and left with 7,539 sources. 
 \textit{AKARI}$-$6dFGS sample includes  4,717 IR galaxies with extinction corrected $b_{j} \leq 16.75 $ within the $0.01 \leq z \leq 0.3$ redshift range.
We do not include galaxies at $z < 0.01$ since they would have  unreliable spectroscopic redshift estimates due to their relatively large  peculiar velocities.

We note that we identify duplicates between the SDSS and 6dFGS samples. 
When a duplicate is identified, if the redshift difference is below 0.002, it is kept in the SDSS sample. 
If the redshift difference is greater than 0.002, the spectra are checked by eye and kept in the SDSS sample if the spectra were reliable. 

\subsubsection{The AKARI$-$2MRS sample}\label{S:2mrs}

The Two Micron All Sky Survey \citep[2MASS][]{Skrutskie2006} completed an all-sky survey in \textit{J}, \textit{H}, and \textit{K$_{s}$} bands. 
The 2MASS Redshift Survey \citep[2MRS][]{Huchra2005,Huchra2012} completed a magnitude-limited (\textit{K$_{s}$} = 11.75 mag) spectroscopic 
survey of 2MASS galaxies over $\sim$91\% of the sky.  
The 2MRS catalog \citep{Huchra2012} presents redshifts for 43,533 galaxies. 
The initial \textit{AKARI} sample is cross-matched with the 2MRS catalog. 
We identified 11,282 IR galaxies in the initial  \textit{AKARI}$-$2MRS sample. 
When the duplicates in the \textit{AKARI}$-$SDSS and \textit{AKARI}$-$6dFGS samples eliminated, the final  \textit{AKARI}$-$2MRS sample includes 6,216 
galaxies within the $z \leq 0.3$ redshift range.

 \section{Results and Discussion} \label{S:results} 
 
 \subsection{The Total Infrared Luminosity Measurements} \label{S:Lir}

We measure the total infrared luminosity of 15,638 galaxies in the \textit{AKARI}$-$SDSS, \textit{AKARI}$-$6dFGS and  \textit{AKARI}$-$2MRS samples by performing SED fitting using the \textsc{lephare}\footnote[7]{http://www.cfht.hawaii.edu/~arnouts/lephare.html} \citep[Photometric Analysis for Redshift Estimations, ][]{Arnouts1999,Ilbert2006}. 
The \textsc{lephare} code finds the best-fitting galaxy template from a given SED library by a $\chi^{2}$ minimisation according to the input photometric magnitudes and redshift. 
For our sample we use the SED library of \citet{Dale2002} which represents the far-IR SEDs of IR galaxies. 
We use the six \textit{AKARI} bands, the four \textit{WISE} bands (when available) with the associated optical photometry ($u$, $g$, $r$, $i$, $z$, $b_{j}$, $K_{s}$) and fix the redshift of each galaxy to the spectroscopic one in the fitting procedure. 
The $k$ corrections are obtained by integrating the filter response functions in the best-fitting SED.
We obtain $L_{IR}$ integrated over 8$-$1000 $\mu$m  with the upper and lower uncertainties of $L_{IR}$ based on the photometric flux uncertainties. 
The $L_{IR}$ measurements are listed in Table \ref{tab:table1}. 
Table \ref{tab:table1} presents AKARI ID's, coordinates, IR and optical photometry, redshift, and IR luminosities of the  \textit{AKARI}$-$SDSS$-$6dFGS$-$2MRS sample.

\begin{landscape}
\begin{table}
           \centering
           \caption{ \textit{AKARI}$-$SDSS$-$6dFGS$-$2MRS sample. The full table in available in the electronic version of the article. 
           NaN value is represented by -99. for photometric values. 
           Columns: (1) \textit{AKARI} ID from the \textit{AKARI}/FIS all-sky survey bright source catalog version 2. 
           (2) and (3) \textit{AKARI} coordinates in the  \textit{AKARI}/FIS all-sky survey bright source catalog version 2. 
            (4) - (7):  \textit{WISE} $W1$, $W2$, $W3$, $W4$ magnitudes and their errors. These are the elliptical aperture magnitudes for the sources with $ext\_flg > 0$. 
            (8) - (13): The \textit{AKARI} flux densities at 9, 18, 65, 90, 140, and 160 $\mu$m, and their associated uncertainties from the \textit{AKARI}/FIS all-sky survey bright source catalog version 2. and 
             \textit{AKARI}/IRC all-sky survey point source catalog version 1.
            (14): Galactic extinction corrected SDSS Petrosian $r$ magnitude. 
            (15): Galactic extinction corrected  $b_{j}$ magnitudes (AB mag) adopted from SuperCOSMOS all-sky galaxy catalogue \citep{Peacock2016}.
            (16): Galactic extinction corrected 2MASS $K_{s}$ magnitude (AB mag) from 2MASS Redshift Survey catalog \citep{Huchra2012}.
            (17): Redshift based on optical spectra. 
            (18): Total IR luminosity between 8 and 1000$\mu$m measured from the SEDs fitting. 
            (19): Best-fitted SED model number from \citet{Dale2002} libary.  
             }
           \label{tab:table1}
           \begin{tabular}{lrrccccccccccccccccc}
           \hline
           \textit{AKARI}         & \textit{AKARI}  & \textit{AKARI}  &  $W1$  & $W2$ & $W3$ &  $W4$ & $F$(9  &$F$(18 & $F$(65 & $F$90 & $F$(140 & $F$(160 &$r$ &$b_{j}$ & $K_{s}$ &$z$ & $\log$  & SED \\
            source name         & R.A.               & Decl.                &             &  &     &                   & $\mu$m)&$\mu$m)&$\mu$m)&$\mu$m)&$\mu$m)&$\mu$m)    &      &             &               & &$(L_{IR}/L_{\sun}$) & model\\
              (AKARI-FIS-V2)   & (J2000)         & (J2000)            &      &  &  &  & & &  &  & &                                    &  & &  & &  &\\
                                           & (deg)         & (deg)            &   (mag)  & (mag) & (mag) & (mag) & (Jy)&(Jy) & (Jy) & (Jy) &(Jy) & (Jy)                                      & (mag) & (mag) & (mag) & &  & \\
           \hline
           0042256$+$144201&10.61&14.70&-99&-99&-99&-99&-99&-99& 9.95$\pm$ 0.48& 9.35$\pm$ 0.07& 8.45$\pm$ 0.13& 7.93$\pm$ 0.17&15.40&-99&-99&0.04& $10.72_{ 0.06}^{ 0.00}$ &64 \\              
           0058485$+$283041&14.70&28.51&-99&-99&-99&-99&-99&-99&-99 & 9.38$\pm$ 0.08& 8.28$\pm$ 0.10& 7.84$\pm$ 0.18&16.62&-99&-99&0.11& $11.70_{ 0.06}^{ 0.02}$ &64 \\ 
           0106247$-$014153 &16.60&-1.70&-99&-99&-99&-99&-99&-99&  9.17$\pm$  0.29& 9.64$\pm$ 0.12& 9.99$\pm$ 0.48& 7.20$\pm$ 0.09&16.61&-99&-99&0.08& $11.26_{ 0.08}^{ 0.01}$ &64 \\                         
           0110267$-$084457  &17.61&-8.75&-99&-99&-99&-99&-99&-99& 11.47$\pm$  1.45& 9.51$\pm$ 0.12&-99 & 7.47$\pm$ 0.11&15.86&-99&-99&0.05& $11.04_{ 0.05}^{ 0.01}$ &64 \\               
           0138528$-$102708 &24.72   &-10.45&-99&-99&-99&-99&-99&-99&  7.14$\pm$  0.06& 6.87$\pm$ 0.01&  7.40$\pm$  0.06& 7.81$\pm$ 0.15&16.13&-99&-99&0.05& $11.75_{ 0.03}^{ 0.02}$ &25 \\                        
           0151223$+$130335  &27.84& 13.06&-99&-99&-99&-99&-99&-99&  9.77$\pm$  0.68& 9.27$\pm$ 0.11&  8.44$\pm$  0.15& 7.83$\pm$ 0.19&14.75&-99&-99&0.06& $11.22_{ 0.06}^{ 0.01}$ &64 \\  
		\hline
	\end{tabular}
\end{table}  
\end{landscape}  

 \subsection{Luminosity Functions} \label{S:lf}

\subsubsection{The $1/V_{max}$ method} \label{S:vmax}

We use $1/V_{max}$ method \citep{Schmidt1968} to compute LF for local IR galaxies. 
We prefer the $1/V_{max}$ method because it derives the LF directly from the data, without any model/parameter assumption on the LF shape. 
The $1/V_{max}$ technique is based on the galaxy number counts within a volume. 
The maximum comoving volume ,$V_{max}= V_{z_{max}}-V_{z_{min}}$, is calculated for each galaxy from the maximum redshift, $z_{max}$, at which it can be still detected in the considered surveys. 
In order to get the $z_{max}$ for the F(90$\mu$m) flux limit, we compute the $k$-correction based on the same SEDs that the $L_{8-1000}$ luminosities were measured. 
We obtain $k$-corrections of the SDSS $r_{petro}$, 6dFGS $b_{j}$ and 2MASS $K_{s}$ magnitudes by using the kcorrect code (v4\_3) of \citet[][]{Blanton2007}. 
Once the $k$-corrections are obtained for each individual galaxy, 
it is moved to the redshift where F(90$\mu$m) and optical magnitudes ($r_{petro}$ or $b_{j}$ or $K_{s}$) reach their limits, respectively.  
We take into account the nominal survey limits that the surveys are complete; for the SDSS extinction corrected $r_{petro} \leq 17.7$, 6dFGS extinction corrected $b_{j} \leq 16.75 $, 
and for 2MRS $K_{s} \leq 11.75$. 
For  \textit{AKARI} we use F(90$\mu$m)=0.5 Jy flux limit where completeness is expected to be close to 80 per cent based on the completeness counts given 
for \textit{AKARI}/FIS all-sky survey bright source catalog version 1 \citep{Goto2011}. 
We use the detection completeness curve \citep[Fig. 8 of ][]{Yamamura2010} to obtain the completeness of each source based on the 90$\mu$m flux density.
If $z_{max}$ exceeds the upper redshift limit of our local IR galaxy sample we set it to 0.3. 
We limit $z_{min}$ to be the minimum redshift limit of the considered sample.

We consider total infrared luminosity  bins between $\log [L_{IR}/L_{\sun}]=$8 and 13, each bin has the size of 0.3 dex. 
In each luminosity bin, the LF is computed as
\begin{equation}\label{Eq:lf}
\Phi(L)=\frac{1}{\Delta L} \sum_{i} \frac{1}{w_{i} \times V_{max,i}}
 \end{equation}
where $V_{max,i}$ is the comoving volume over which the $i$th galaxy could be detected, $\Delta L$ is the size of each luminosity bin and 
$w_{i}$ is the completeness correction factor for the ith galaxy. 
This accounts for the IR detection incompleteness and the sky coverage correction, $w$=(1/completeness)*(all sky area/survey area). 
We adopt the effective survey areas estimated by the 6dFGS DR3 and 2MRS data releases \citep{Jones2009,Huchra2012}. 
The used effective survey areas of 6dFGS and 2MRS surveys are 13,572 and 37,000 square degrees, respectively. 
In order to estimate the SDSS DR 13 legacy spectroscopic sky coverage we produce a map of the \textit{specObj} catalog {\it programname = legacy} sources 
with HEALPIX\footnote[4]{\textsc{healpix} is hierarchical equal area isolatitude pixelization of a sphere, see https://healpix.jpl.nasa.gov for details.} \citep{Gorski2005}. 
We estimate the SDSS DR 13 spectroscopic effective survey area as  9,219 square degrees. 

The uncertainties on the $1/V_{max}$ data points take into account only Poisson errors (depends on the number of sources); however 
the errors in the photometric fluxes propagate into uncertainties in the measured LF. 
We perform a Monte Carlo simulation in order to analyse this effect. 
Namely, we obtain the LF on 100 different mock galaxy catalogs following the same procedure as for the original sample. 
Each of the mock catalogs contains the same galaxies as the original sample, 
but we generate random fluxes from a Gaussian distribution around the measured flux with  
a dispersion of the measured uncertainties. 
As a result of the Monte Carlo simulation we obtain a Gaussian distribution of $1/V_{max}$ data points in each luminosity bin; 
we assign the standard deviation of the distribution as the Monte Carlo uncertainty. 
Then in our original $1/V_{max}$ results, the total uncertainty in each luminosity bin is the quadratic sum of the Poissonian error and the Monte Carlo uncertainty.

We derive LF for the  \textit{AKARI}$-$SDSS, \textit{AKARI}$-$6dFGS and the combined \textit{AKARI}$-$SDSS$-$6dFGS$-$2MRS samples separately. 
We list the $1/V_{max}$ data points in Table \ref{tab:table2}.

\begin{table*}
           \centering
           \caption{ The IR LF of \textit{AKARI}$-$SDSS,  \textit{AKARI}$-$6dFGS and  \textit{AKARI}$-$SDSS$-$6dFGS$-$2MRS galaxies obtained with the $1/V_{max}$ method. 
           N is the number of sources in each luminosity bin.}
           \label{tab:table2}
           \begin{tabular}{rcrcrcr}
           \hline
           $\log (L_{IR}/L_{\sun})$  &  $\phi$ (Mpc$^{-3}$ dex$^{-1}$) &N & $\phi$ (Mpc$^{-3}$ dex$^{-1}$) &N&  $\phi$ (Mpc$^{-3}$ dex$^{-1}$) &N\\
                                                & \textit{AKARI}$-$SDSS           & & \textit{AKARI}$-$6dFGS         &  & \textit{AKARI}$-$SDSS$-$6dFGS$-$2MRS &\\
           \hline
 8.15&0.000  & 0 &0.000                                                 & 0 & 9.763 $\times\,10^{-3} \pm$ 3.077 $\times\,10^{-3}$ &18 \\
 8.45&0.000  & 0 &2.675 $\times\,10^{-5} \pm$ 3.322 $10^{-5}$ & 1 & 1.328 $\times\,10^{-2} \pm$ 2.985 $\times\,10^{-3}$ & 45 \\
 8.75&0.000  & 0 &8.904 $\times\,10^{-6} \pm$ 1.171 $10^{-5}$ & 1 & 1.134 $\times\,10^{-2} \pm$ 1.592 $\times\,10^{-3}$ & 103 \\
 9.05&0.000  & 0 &5.517 $\times\,10^{-4} \pm$ 2.682 $10^{-4}$ & 12 &  1.051 $\times\,10^{-2} \pm $ 1.235 $\times\,10^{-3}$ & 195 \\
 9.35&9.600 $\times\,10^{-5} \pm $ 1.329 $\times\,10^{-4}$ & 2 &2.306 $\times\,10^{-3} \pm$ 6.352 $\times\,10^{-4}$ & 49 &  7.127 $\times\,10^{-3} \pm$ 5.115 $\times\,10^{-4}$ & 412 \\
 9.65&6.675 $\times\,10^{-4} \pm $ 2.114 $\times\,10^{-4}$ & 30 &3.806 $\times\,10^{-3} \pm$ 4.446 $\times\,10^{-4}$ & 239 & 7.056 $\times\,10^{-3} \pm$ 3.256 $\times\,10^{-4}$ & 982\\
 9.95&4.170 $\times\,10^{-3} \pm $ 4.392 $\times\,10^{-4}$ & 311 & 3.096 $\times\,10^{-3} \pm$ 2.605 $\times\,10^{-4}$  &549     & 6.763 $\times\,10^{-3} \pm$  2.188 $\times\,10^{-4}$ & 2088 \\
 10.25&3.472 $\times\,10^{-3} \pm $ 2.288 $\times\,10^{-4}$ & 946 &1.635 $\times\,10^{-3} \pm$ 7.413 $\times\,10^{-5}$  &982      & 4.933 $\times\,10^{-3} \pm$ 1.217 $\times\,10^{-4}$ & 3466\\
10.55&1.358 $\times\,10^{-3} \pm $ 6.358 $\times\,10^{-5}$ & 1284 &9.023 $\times\,10^{-4} \pm$ 6.711 $\times\,10^{-5}$   & 1281    & 2.344 $\times\,10^{-3} \pm$ 6.652 $\times\,10^{-5}$  & 3798 \\
10.85&3.316 $\times\,10^{-4} \pm $ 1.772 $\times\,10^{-5}$ & 913 &2.571 $\times\,10^{-4} \pm$ 1.511 $\times\,10^{-5}$     & 916   & 6.231 $\times\,10^{-4} \pm$ 2.010 $\times\,10^{-5}$ & 2387 \\
11.15&1.010 $\times\,10^{-4} \pm $ 7.081 $\times\,10^{-6}$ & 617 &8.112 $\times\,10^{-5} \pm$ 8.744 $\times\,10^{-6}$      & 479   & 1.842 $\times\,10^{-4} \pm$ 1.014 $\times\,10^{-6}$ & 1249 \\
11.45&3.270 $\times\,10^{-5} \pm $ 2.945 $\times\,10^{-6}$ & 372 &1.645 $\times\,10^{-5} \pm$ 2.281 $\times\,10^{-6}$    &147    & 4.950 $\times\,10^{-5} \pm$  3.572 $\times\,10^{-6}$ & 562 \\
11.75&6.400 $\times\,10^{-6} \pm $ 8.125 $\times\,10^{-7}$ & 162 &3.137 $\times\,10^{-6} \pm$ 5.734 $\times\,10^{-7}$    &50    &  9.600 $\times\,10^{-6} \pm$ 1.014 $\times\,10^{-7}$ & 220 \\
12.05&1.000 $\times\,10^{-6} \pm $ 2.864 $\times\,10^{-7}$ & 46 &3.103 $\times\,10^{-7} \pm$ 1.283 $\times\,10^{-7}$         &7  & 1.300 $\times\,10^{-6} \pm$ 2.610 $\times\,10^{-7}$ & 55 \\
12.35&1.000 $\times\,10^{-7} \pm $ 3.475 $\times\,10^{-8}$ & 15 &1.208 $\times\,10^{-7} \pm$ 6.889 $\times\,10^{-8}$          &4 & 2.000 $\times\,10^{-7} \pm$  7.709 $\times\,10^{-8}$ & 20 \\
12.65&1.000 $\times\,10^{-7} \pm $ 2.314 $\times\,10^{-8}$  & 7 & 0.000                                                                                 &0 & 1.000 $\times\,10^{-7} \pm$ 2.322 $\times\,10^{-8}$ & 8 \\
		\hline
	\end{tabular}
\end{table*}

\subsubsection{Luminosity function of AKARI$-$SDSS galaxies } \label{S:sdsslf}

\begin{figure}
\begin{center}$
\begin{array}{c}
\includegraphics[]{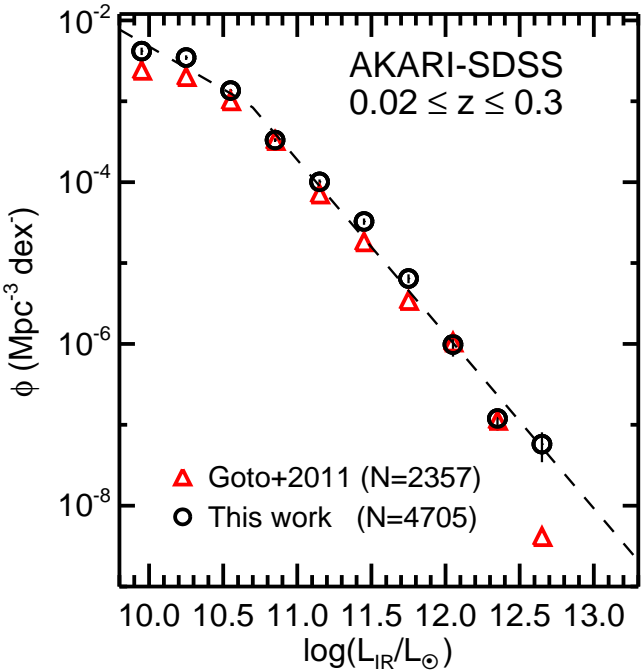}\\ 
\end{array}$
\end{center}
\caption{The IR LF of 5,388 \textit{AKARI}$-$SDSS galaxies. The dashed line is the best-fitting double power law. 
The black open circles are the $1/V_{max}$ data points obtained in this work. 
The uncertainties are the sum of Poisson errors and the errors computed using 100 Monte Carlo simulations. 
The triangles are the results for 2,357 \textit{AKARI}$-$SDSS galaxies obtained by \citet{Goto2011}. } 
\label{fig:fig1}
\end{figure}

The IR LF of 4,705 \textit{AKARI}$-$SDSS galaxies computed with the $1/V_{max}$ method is shown in Fig. \ref{fig:fig1} (open circles). 
The 90$\mu$m flux limit of  \textit{AKARI} at the median redshift of 0.036 is $\log [L_{IR}/L_{\sun}] \sim$10.0 and it imposes the completeness region of our sample. 
Therefore, we fit the $1/V_{max}$ data points starting from  $\log [L_{IR}/L_{\sun}]=$10.0. 
We simply fit the $1/V_{max}$ data points using a double power law \citep{Babbedge2006} as follows: 
\begin{equation}\label{Eq:doublepl1}
\phi(L)dL/L^{*}= \phi^{*} \bigg(\frac{L}{L^{*}}\bigg)^{1-\alpha} dL/L^{*}, L < L^{*}
 \end{equation}
and 
\begin{equation}\label{Eq:doublepl2}
\phi(L)dL/L^{*}= \phi^{*} \bigg(\frac{L}{L^{*}}\bigg)^{1-\beta} dL/L^{*}, L > L^{*},
 \end{equation}
where, $\phi$ is the normalisation factor in Mpc$^{-3}$, $L^{*}$ is the characteristic luminosity (in units of $L_{\sun}$) where the break between the faint and bright regions, $\alpha$ is the faint-end slope and $\beta$ is the bright-end slope. 
The obtained best-fitting parameters for the IR LF of  \textit{AKARI}$-$SDSS sample are $\phi=89 \pm 3 \times\, 10^{-5}$, $L^{*}=4.85 \pm 0.01 \times 10^{10}$, $\alpha=2.06  \pm 0.05$, $\beta=3.1 \pm 0.03$. 
We list the best-fitting parameters in Table \ref{tab:table3}.
The best-fitting double power law is shown as the dashed line in Fig. \ref{fig:fig1}. 
 
\citet{Goto2011} constructed LF of 2,357 IR \textit{AKARI}$-$SDSS DR 7 galaxies. 
Our sample of SDSS DR 13 galaxies is almost twice as large as their \textit{AKARI}$-$SDSS sample. 
Their results are shown as the red triangles in Fig.  \ref{fig:fig1}. 
Our counts are slightly higher compared to that of \citet{Goto2011} in the faint-end . 
This difference is probably because we have a deeper, more complete \textit{AKARI} catalog (ver2) that recovers the missing sources in \citet{Goto2011}, which used ver. 1. 
In fact the number of \textit{AKARI}  sources above the detection limit adopted in \citet{Goto2011} is by a factor of 1.6 larger in ver. 2 than in ver. 1. 
This is one of the major causes of the difference in resulting LFs between these studies. 
In the bright-end part, our $1/V_{max}$ measurements in the $\log [L_{IR}/L_{\sun}]=$12.65 bin is factor of 14 larger. 
This may explain the steeper bright-end slope ($\beta=3.54 \pm 0.09$) and the larger IR LF break ($L^{*}$) \citet{Goto2011} obtain. 
Our faint-end slope is consistent with the one given by \citet{Goto2011}, within the two-sigma uncertainty level. 
The doubled sample size improves the underestimated high luminosity end measurement significantly. 

 \subsubsection{Luminosity function of AKARI-6dFGS galaxies } \label{S:6dflf}
\begin{figure} 
\begin{center}$
\begin{array}{c}
\includegraphics[]{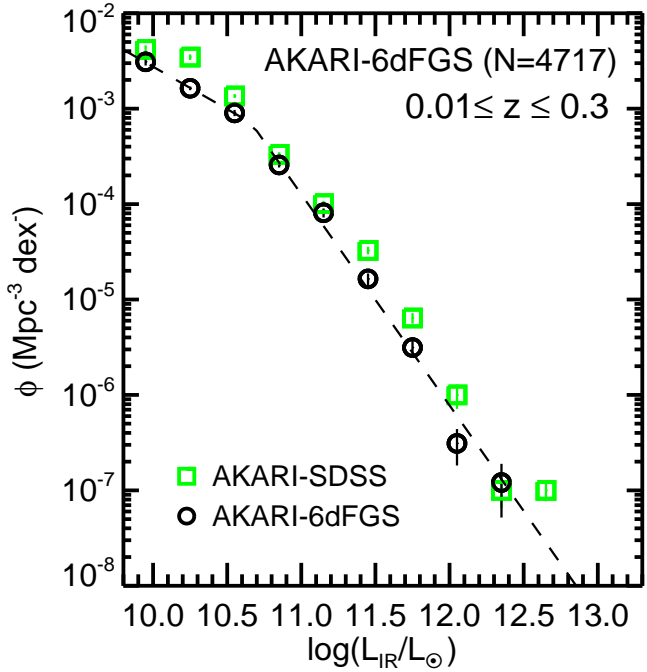}\\ 
\end{array}$
\end{center}
\caption{The IR LF of 4,717  \textit{AKARI}$-$6dFGS galaxies. The dashed line shows the best-fitting double power law. 
The black open circles are the $1/V_{max}$ measurements with errors obtained from 100 Monte Carlo simulations added by the Poisson error.). 
The squares are the results for the \textit{AKARI}$-$SDSS sample obtained in this work.} 
\label{fig:fig2}
\end{figure}

 We show the LF of 4,717  \textit{AKARI}$-$6dFGS galaxies in Fig. \ref{fig:fig2}. 
 We fit the LF using the double-power law as expressed in Equations \ref{Eq:doublepl1} and  \ref{Eq:doublepl2} using a $\chi^{2}$ minimization. 
 The best-fitting values are given in Table \ref{tab:table2}.
 
The green squares show the $1/V_{max}$ measurements of the \textit{AKARI}$-$SDSS galaxies (\S \ref{S:sdsslf}). 
In general, $1/V_{max}$ data points measured for the SDSS sample are factor of 2 higher than that of the 6dFGS sample. 
This difference can be related to the different effective survey areas of the SDSS and 6dFGS samples. 
The 6dFGS sample has a factor of 1.5 larger effective area that results in a larger volume and a lower normalization. 
The number of galaxies in each luminosity bin is different for each sample; at the highest luminosity bin ($\log [L_{IR}/L_{\sun}]=$12.95) there are no galaxies in the 6dFGS. 
Therefore we expect to have disagreement in some of the best-fitting values. 
While the faint- and bright-end slopes agree within four-sigma uncertainty level; the IR LF break of the 6dFGS sample is lower compared to the SDSS sample. 

 \subsubsection{Luminosity function of AKARI-SDSS-6dFGS-2MRS galaxies } \label{S:sdss6dflf}
\begin{figure} 
\begin{center}$
\begin{array}{c}
\includegraphics[]{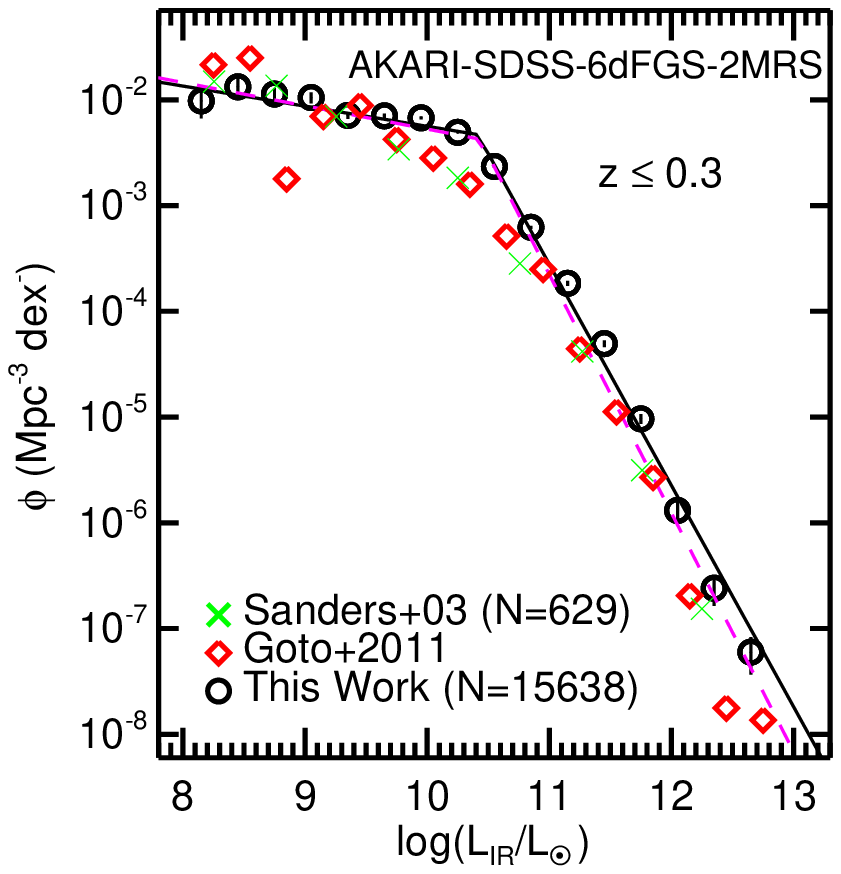}\\ 
\end{array}$
\end{center}
\caption{The IR LF of 15,638  \textit{AKARI}$-$SDSS$-$6dFGS$-$2MRS }  galaxies (open circles). 
The best-fitting double power law is shown as solid line.
For comparison the total IR LF derived from the \textit{IRAS} RBGS is shown \citep[crosses][]{Sanders2003}.
The red diamonds are the $1/V_{max}$ data points of the RBGS sample adopted from \citet{Goto2011a}. 
The dashed magenta line is the best-fitting double power law when the RBGS data are included in the fit. 
\label{fig:fig3}
\end{figure}

 The $1/V_{max}$ LF of the combined 15,638  \textit{AKARI}$-$SDSS$-$6dFGS$-$2MRS galaxies is shown in Fig. \ref{fig:fig3}. 
 The best-fitting double power law is shown as the black solid line. 
 SDSS and 6dFGS surveys are incomplete at bright magnitudes, due to the incompleteness below $\log [L_{IR}/L_{\sun}]=$10.0 we cannot measure the faint-end of the IR LF. 
 In order to extend the LF to lower luminosity bins with a complete sample of local galaxies we use the 2MRS sample.
 The addition of the 2MRS survey, especially the  lower redshift ($z < 0.01$) galaxies in the sample allow us to probe the faint-end. 
 However, we caution that below $\log [L_{IR}/L_{\sun}]=$9.65 the completeness is not 100 percent and therefore the obtained LF represents a lower limit. 
 Not surprisingly, the bright-end slopes of the combined sample agrees with the SDSS and 6dFGS samples within three-sigma uncertainty. 
 Compared to these two samples, for the combined sample the IR LF break is at a lower luminosity of $L^{*}=2.60 \times 10^{10}$.

 We compare our result with the IR LF derived by \citet{Sanders2003} from a complete sample of 629 60 $\mu$m 
 selected \textit{IRAS} galaxies (green crosses). This is the \textit{IRAS} Revised Bright Galaxy Sample (RBGS). 
 We also show the $1/V_{max}$ data (red diamonds) of  the \textit{IRAS} RBGS from \citet{Goto2011a}. 
 Our results are in agreement with the results of  \citet{Sanders2003} and \citet{Goto2011a}. 
 \citet{Goto2011a} follow a similar procedure to obtain $L_{IR}$ and the $1/V_{max}$ measurements, also they include 160 $\mu$m band in estimating $L_{IR}$.
 When the RBGS data are added, we obtain a very similar fit (dashed magenta line in Fig. \ref{fig:fig3}). 
 Compared to \citet{Goto2011a} we obtain flatter faint- and bright-end slopes and a lower $L^{*}$. 
 
 This shows that by increasing the number of local IR galaxies more than 20 times, we obtain the most accurate LF which is consistent with the results of previous studies \citep{Sanders2003,Goto2011a}. 
 The most significant effect of the larger sample size is the much smaller statistical uncertainties (i.e., root mean square error is $(\sum V_{max}^{-2})^{0.5}$) in each luminosity bin. 
 However, this does not necessarily decrease the uncertainties of the best-fitting parameters. 
 
 We note that due to the lack of complete (optical or IR) datasets that could provide AGN and star forming galaxy (SFG) separation for the combined 15,638  \textit{AKARI}$-$SDSS$-$6dFGS$-$2MRS galaxies, 
 we do not attempt to separate individual galaxies into AGN/SFG. Therefore, the LFs presented in this work includes the contribution from AGN. We also note that in far-IR, the contribution from the warm dust of AGN is expected to be small, and the emission is expected to be dominated by star-formation activity.
 
\begin{table*}
           \centering
           \caption{ Best-fitting double power-law IR LF parameters. Columns: (1) Sample name. (2) Number of galaxies in the sample. 
           (3)Break luminosity of the double power law LF. (4) Faint-end slope of the double power law LF. (5) Bright-end slope of the double power law LF. 
	}
           \label{tab:table3}
           \begin{tabular}{lcccccr}
           \hline
           Sample & N & $L_{IR}^{*}$ & $ \phi ^{*}$& $\alpha$ & $ \beta$ & $\Omega_{IR}$\\
                        &    & [$L_{\sun}$] & [Mpc$^{-3}$ dex$^{-1}$] &  & &  [$L_{\sun}$ Mpc$^{-3}]$\\
           (1) & (2) & (3) & (4)  & (5) & (6) & (7) \\
           \hline
 \textit{AKARI}$-$SDSS                 &  4705  & 4.85 $\pm$ 0.01 $\times\, 10^{10}$ & 89 $\pm$ 3 $\times\, 10^{-5}$ & 2.06 $\pm$ 0.05 & 3.16 $\pm$ 0.03  & 10.15 $\pm$ 2.78 $\times\, 10^{8}$\\ 
\textit{AKARI}$-$6dFGS                &  4717  & 4.60 $\pm$ 0.01 $\times\, 10^{10}$  &70 $\pm$ 3 $\times\, 10^{-5}$ & 1.89 $\pm$ 0.06 & 3.21 $\pm$ 0.05  & 1.32 $\pm$ 0.11 $\times\, 10^{8}$\\
\textit{AKARI}$-$SDSS$-$6dFGS$-$2MRS & 15638 & 2.60 $\pm$ 0.05 $\times\, 10^{10}$ & 473 $\pm$ 13 $\times\, 10^{-5}$ & 1.22 $\pm$ 0.02 & 3.06 $\pm$ 0.02  & 1.19 $\pm$ 0.05 $\times\, 10^{8}$\\ 
 \textit{AKARI}$-$SDSS$-$6dFGS$-$2MRS+RBGS & 16237 & 2.71 $\pm$ 0.05 $\times\, 10^{10}$ &424 $\pm$ 12 $ \times\, 10^{-5}$ & 1.22 $\pm$ 0.02 & 3.26 $\pm$ 0.02  & 1.04 $\pm$ 0.06 $\times\, 10^{8}$\\ 
		\hline
	\end{tabular}
\end{table*}

 \subsection{Total Infrared Luminosity Density} \label{S:LD}
 
 We integrate the measured LFs to estimate the total IR luminosity density, $\Omega_{IR}$. 
 For each sample we integrate the best-fitting double power law. 
 We list the obtained total luminosity densities in Table \ref{tab:table3}. 
 The total luminosity density of the combined sample of 15,638 IR galaxies is $\Omega_{IR}= 1.19\pm0.05 \times 10^{8} L_{\sun}$ Mpc$^{-3}$. 
  In order to estimate the uncertainty we add and subtract the 1$\sigma$ error from the best-fitted double power law, we take the largest difference as the error. 
Only a small fraction of $\Omega_{IR}$ is produced by luminous IR galaxies and ultra luminous IR galaxies; 9.3 per cent and 0.9 per cent, respectively. 
This is consistent with the results of previous studies \citep[e.g.,][]{Soifer1991,Goto2011} that quantified the contribution of U/LIRGs to $\Omega_{IR}$ as small.

\section{Conclusions}\label{S:conc}
In our aim to derive the most accurate local IR LF, we have cross-matched the \textit{AKARI} all-sky survey with the SDSS, 6dFGS and 2MRS to find 15,638 IR galaxies with spectroscopic redshifts. 
We have measured $L_{IR}$ by SED fitting based on the six \textit{AKARI}, the four \textit{WISE} IR photometry bands. 
Especially, the far-IR photometry that cover the peak of the dust emission SED allow us to have accurate $L_{IR}$ measurements compared to the ones based on a bolometric conversion factor.
Our main conclusions are: 

\begin{enumerate}

\item
We obtain the most accurate local (median $z=0.027$) IR LF for the largest sample studied so far. 
Our local IR LF will be a reliable benchmark for future investigations of total IR LF evolution at higher redshifts. 

\item
Compared to the previous study of \citet{Goto2011}, we have doubled the \textit{AKARI}$-$SDSS sample size. 
Our LF with better high luminosity-end measurements, is consistent with that of \citet{Goto2011}. 

\item
Our analysis of the LF of different samples agree with each other. 
Particularly we measure a similar bright-end slope for different samples. 
With much greater precision they are also consistent with the previous measurements.   

\item
For the combined sample of 15,638 IR galaxies, we compute the local IR luminosity density as $\Omega_{IR}= 1.19\pm0.05 \times 10^{8} L_{\sun}$ Mpc$^{-3}$.  
In the local Universe, the contribution of LIRGs and ULIRGs to $\Omega_{IR}$ is very small; 9.3 per cent and 0.9 per cent, respectively.

\end{enumerate}

\subsection*{Acknowledgments} 
We thank Mattia Vaccari for many insightful comments. 
TG acknowledges the support by the Ministry of Science and Technology of Taiwan through grant NSC 103-2112-M-007-002-MY3, and 105-2112-M-007-003-MY3.
This research is based on observations with AKARI, a JAXA project with the participation of ESA. 
This research has made use of the ASPIC database, operated at CeSAM/LAM, Marseille, France.
Funding for the Sloan Digital Sky Survey IV has been provided by
the Alfred P. Sloan Foundation, the U.S. Department of Energy Office of
Science, and the Participating Institutions. SDSS-IV acknowledges
support and resources from the Center for High-Performance Computing at
the University of Utah. The SDSS web site is www.sdss.org.
SDSS-IV is managed by the Astrophysical Research Consortium for the 
Participating Institutions of the SDSS Collaboration including the 
Brazilian Participation Group, the Carnegie Institution for Science, 
Carnegie Mellon University, the Chilean Participation Group, the French Participation Group, Harvard-Smithsonian Center for Astrophysics, 
Instituto de Astrof\'isica de Canarias, The Johns Hopkins University, 
Kavli Institute for the Physics and Mathematics of the Universe (IPMU) / 
University of Tokyo, Lawrence Berkeley National Laboratory, 
Leibniz Institut f\"ur Astrophysik Potsdam (AIP),  
Max-Planck-Institut f\"ur Astronomie (MPIA Heidelberg), 
Max-Planck-Institut f\"ur Astrophysik (MPA Garching), 
Max-Planck-Institut f\"ur Extraterrestrische Physik (MPE), 
National Astronomical Observatories of China, New Mexico State University, 
New York University, University of Notre Dame, 
Observat\'ario Nacional / MCTI, The Ohio State University, 
Pennsylvania State University, Shanghai Astronomical Observatory, 
United Kingdom Participation Group,
Universidad Nacional Aut\'onoma de M\'exico, University of Arizona, 
University of Colorado Boulder, University of Oxford, University of Portsmouth, 
University of Utah, University of Virginia, University of Washington, University of Wisconsin, 
Vanderbilt University, and Yale University.


\bibliographystyle{mnras}
\bibliography{lfpaper}




\bsp	
\label{lastpage}
\end{document}